\documentclass[titlepage,twoside,12pt]{article}
\usepackage{amssymb}
\usepackage{amsfonts}
\begin{document}

{\bf \Large Quantum Foundations : \\ \\ Is Probability Ontological ?} \\ \\

Elem\'{e}r E Rosinger \\ \\
Department of Mathematics \\
and Applied Mathematics \\
University of Pretoria \\
Pretoria \\
0002 South Africa \\
eerosinger@hotmail.com \\ \\

{\bf Abstract} \\

It is argued that the Copenhagen Interpretation of Quantum Mechanics, founded ontologically on
the concept of probability, may be questionable in view of the fact that within Probability
Theory itself the ontological status of the concept of probability has always been, and is
still under discussion. \\ \\

{\bf 1. Two Issues in Probability Theory} \\

Two long unresolved - and in fact, seldom considered - issues have for more than two centuries by 
now accompanied modern Probability Theory. \\
The first one we shall only mention as an example of how fundamental aspects can be - and in
fact, are - overlooked. The second one, which has not been completely overlooked, has
nevertheless found itself side lined for longer, even if it has obvious major implications in
Quantum Foundations. \\

As for the more general background on the various, and often severely conflicting views and
interpretation of the basic concepts or methods of Probability Theory, a recent survey can be
found in [1]. \\

One of the two issues mentioned above, namely, the issue of {\it redundancies}, [2], seems not
to have been the object of any wider awareness, although it has led to considerable technical
difficulties in, among others, stochastic processes with continuous time. And as it happens,
this issue has not been overcome even by the nonstandard approach to probabilities,
specifically, by the introduction of Loeb measures and integration. \\

The second issue, which is known at least in principle, seems nevertheless similarly to be
outside of the realms of general awareness, and certainly, of a more active pursuit in its
possible implications in a large variety of applications of Probability Theory. This issue
concerns whether the axioms of Probability Theory, as given for instance in their formulation
due to Kolmogorov, happen to have a deeper status than mere {\it epistemic convenience}. \\ \\

{\bf 2. Deterministic, versus Non-Deterministic Phenomena} \\

Classical Physics, as well as Special and General Relativity, typically deal with {\it
deterministic} phenomena, such as for instance those related to gravitation, electro-magnetism,
and so on. \\
What is important to note here is that, usually in such situations, the classification
"deterministic" is not merely an epistemic choice of convenience, but it is rather seen as an
essential ontological feature of the respective phenomena. \\

On the other hand, and not only in the realms of Physics, there are phenomena which obviously
cannot be considered deterministic, be it from an ontological, or for that matter, epistemic point of view. \\
In this regard, until the 1960s, there has been one single subclass of non-deterministic
phenomena specifically identified and studied as such, namely, the subclass composed of those
phenomena which were considered to be {\it probabilistic}. And clearly, even if not always
explicitly expressed, the subclass of probabilistic phenomena was not - and in fact, could not
be - identified with the whole class of non-deterministic ones. \\

However, since the work of Zadeh in the 1960s, a second subclass of non-deterministic
phenomena, namely, those called {\it fuzzy}, has been pointed out, even if not in as clear and
rigorous mathematical formulation as that of the probabilistic ones. And needless to say,
there is no claim that these two classes, namely, the probabilistic and the fuzzy one, may
have a significant overlap. Similarly, there is no - and there could not possibly be - any
claim that these two classes exhaust all the non-deterministic phenomena. \\

Not much later, in the 1970s, with the work of Feigenbaum, a further subclass of phenomena,
namely, the {\it chaotic} ones got singled out. Here however, a somewhat surprising turn of
events happened, as the chaotic phenomena studied ever since have typically been those that
may be seen as "deterministic chaos", that is, chaotic phenomena produced - rather
surprisingly - by certain deterministic nonlinear systems. \\ \\

{\bf 3. Ontological, or merely Epistemic ?} \\

Being here mostly interested in the issue of the ontological, or on the contrary, merely
epistemic status of modern Probability Theory, we shall only consider the following division
in classes of phenomena :

\begin{itemize}

\item deterministic

\item non-deterministic

\begin{itemize}

\item probabilistic

\item fuzzy

\item the rest

\end{itemize}

\end{itemize}

Several remarkable facts, seldom, if at all of a wider concern, are worth mentioning about
these classes. \\

First perhaps is the fact that the class of non-deterministic phenomena is simply defined by a
mere {\it negation}. And as it is a well known elementary fact of Logic, such a definition by
negation is hardly appropriate, since it opens up all the possibilities other than specified
by the negated one, [4-6]. This is therefore a possible major source of the difficulties which
may lurk at the bottom of all subsequent dealings with the resulting concept of
non-deterministic phenomena. In particular, this definition by mere negation may negatively
affect the attempts to set up definitions for the subclasses of probabilistic, fuzzy or
chaotic phenomena, not to mention other possible non-deterministic subclasses. \\

And one of such important negative effects may be in the fact that, although we tend to see
the concept of "deterministic" as having an essentially ontological status, the definition by
a mere negation of the concept of "non-deterministic" may seriously weaken any hope for a
similar status of that concept, and consequently, of its particular cases of concepts, such as
"probabilistic", "fuzzy" "chaotic", and so on. \\

Focusing now on the concept of "probabilistic", it should be recalled that, back in the 1930s,
soon after the setting up of the Kolmogorov model, De Finetti pointed out that, within such a
model there are most serious issues as to the ontological status of the concept of probability.
And in a somewhat shocking and provocative manner, De Finetti stated, [3], that "probability
does not exist!" ... \\

As it happens, such a view is but a part of a larger trend or school of thought, called
usually {\it Subjective Probability}, [1], a trend which incorporates the much older Bayesian
subjectivity which is still in use in certain circles, as well as a number of other ones. And
the main claim of that view is that the concept of probability is {\it lacking} any ontological
status, being instead a mere {\it epistemic} choice of convenience, therefore subjective as
such. \\

Needless to say, in the case of the fuzzy subclass of non-deterministic phenomena, the
ontological status of that concept appears to have even a lesser likelihood than that of the
concept of probability. \\ \\

{\bf 4. Should Quanta Be Founded on Ontologically Uncertain
\hspace*{0.6cm} Probabilities ?} \\

Starting with the mid 1920s, when the modern version of Quantum Mechanics replaced the
original one introduced in 1900 by Planck, and developed by Einstein, Bohr and a few others,
it appeared to be a matter of a significant pride to Bohr, Heisenberg and Born to have
originated such a totally unprecedented new theory of Physics, one which in its rather
incredibly counterintuitive novelty went far beyond even of Einstein's General Relativity. And
in the view of their originators, a view which was to become the so called Copenhagen
Interpretation, a major, if not in fact, by far the most major novelty was what they
considered to be the inevitable and irreducible involvement of randomness or probability in
quantum phenomena. \\
This position of the founders of modern Quantum Mechanics was precisely that which created the
extreme separation between them, and on the other hand, a few others, among them at the time
Einstein, Schr\"{o}dinger, De Broglie, and later, Bohm and Bell. In this regard, the often
cited maxim of Einstein that "God does not play dice" expresses quite clearly the total
opposition between the respective two views. \\

In other words, the founders of modern Quantum Mechanics postulated nothing short of an {\it
ontological} status for probabilities in quantum phenomena. On the other hand, Einstein and
others were only willing to accept for probabilities a mere {\it epistemic} status. In
particular, in Einstein's view, the wave function, and the corresponding probability amplitude,
were only describing an ensemble of quantum particles prepared in the same way, and by no
means one single individual such particle. Thus Einstein's view that Quantum Mechanics, as
advocated by the Copenhagen Interpretation, had to be incomplete, a view famously presented in
the 1935 celebrated EPR paper. \\

But now, with hindsight, the following issue arises :

\begin{itemize}

\item Quantum Mechanics, in its Copenhagen Interpretation, is founded on the ontological
status played in that theory by the concept of probability. On the other hand, that
ontological status of the concept of probability has for long been questioned in Probability
Theory. Therefore, should, or for that matter, can Quantum Mechanics have an ontology which
insists on probabilities ?

\end{itemize}

The fact that the founders of modern Quantum Mechanics did not seem to consider the above
issue is easy to explain. As most of those who apply Probability Theory, they were not much
concerned with the inner affairs of that theory. Not to mention that the work of De Finetti
only started to appear a decade later, in the 1930s. \\

What may, however, be less easy to explain is why even today no concern is shown related to
the mentioned possibly ill-founded ontology of Quantum Mechanics in its Copenhagen
Interpretation. \\ \\

{\bf 5. Infinity Again Causes Trouble ...} \\

One of the basic problems - even if hardly at all known within wider circles of mathematicians or physicists - which keep troubling Probability Theory is the inevitable and essential involvement of {\it infinity}. \\

In this regard, it is worth recalling that Euclidean Geometry has for more than two millennia - that is, until Bolyai and Lobachevski introduced in the early 1800s non-Euclidean Geometry - been also troubled by infinity. Indeed, Euclid, with so many others after him, firmly believed that postulates, or what we call nowadays by the name of axiomxs, must be self-evident, this being the only basis for their acceptance. However, his Fifth Postulate on parallel lines was not only clearly more complex in its formulation than the other postulates, but on top of that, it was also the only one which involved infinity, and as such, it was not within the realms of direct empirical verification. \\
The effect was that for more than two millennia, it got singled out as being less self-evident than the other postulates. And as a result, the general attitude has been to question its status of being a postulate. \\
What happened, however, was that such a questioning could for long only make those involved aware of one single logical option. Namely, if the fifth postulate was not in fact a postulate, then it had to be a consequence of the other postulates, therefore, it could be proved based on the other postulates. \\
As we know, such an approach proved to be wrong, and a second logical possibility turned out to exist, namely that the
fifth postulate was actually {\it independent} of the other ones. And then, one could build a Geometry in which the {\it negation} of the fifth postulate would hold, together with the other axioms of Euclidean Geometry. \\
Bolyai and Lobachevski chose one of the two possible negations, namely that there are many different parallel lines to a given line which pass through a point outside of that line. \\

Now, in Probability Theory infinity appears also inevitably, and does so in several not unrelated ways. Here, for the sake of brevity, we mention two of them. Further details can be found in [1]. \\

An important feature of both of these approaches is that they consider probability as being {\it ontological}, and not merely epistemic. Or in other words, they consider probability to exist as a reality in the world, and not only as a construct in our mind. \\ 

The {\it frequency} based view of probability is indeed most natural in case of a finite number of possible events. However, when attempted to extrapolate it to an {\it infinite} number of events, one faces difficulties which in their foundational aspects are not unlike those faced with the Fifth Postulate of Euclidean Geometry. \\

The {\it propensity} based view of probability seems also rather natural, even if somewhat more involved and subtle than the frequency based one. And similarly, it involves {\it infinity} inevitably and in an essential way. \\

In view of the above, what may indeed turn out - regardless of anything else - to be simply amusing about the Copenhagen Interpretation is the apparent total lack of awareness of its proponents and supporters of the fact that they are inevitably involving themselves in dealing with infinity, and doing so in ways Physics, or for that matter, physicists have never done it before. \\

Indeed, in all other branches of Physics, infinity only appears as a possible {\it quantity} of one or another of the physical entities. And the definition of such entities does not at all involve infinity. Also, empirically one can deal with such entities without having to get involved with infinity, except for special cases which are outside of by far most of the usual encounters with Physics, be they theoretical, experimental, or applicative. \\

On the other hand, according to the Copenhagen Interpretation, which sees probability as ontological in the realms of quanta, the inevitable and essential involvement of infinity is there, as mentioned above, and it is there in completely new - yet consciously hardy known - ways in Physics ...


\begin{thebibliography}{99}

\bibitem{} Interpretations of Probability. \\
http://plato.stanford.edu/archives/sum2003/entries/probability-interpret/

\bibitem{} Rosinger E E : Redundancies : An Omission in Probability \\
Theory ? arXiv:math.GM/0608141

\bibitem{} Nau R F : De Finetti was right : probability does not exist. Theory and Decisions,
Vol. 51, 2001, 89-124

\bibitem{} Rosinger E E : Further de-Entangling Entanglement. \\ arXiv:physics/0701246

\bibitem{} Rosinger E E : A General Scheme of Entanglement. \\ arXiv:physics/0701116

\bibitem{} Rosinger E E : Theories of Physics and Impossibilities. \\ arXiv:physics/0512203

\end{thebibliography}
\end{document}